# The histone octamer influences the wrapping direction of DNA on it: Brownian dynamics simulation of the nucleosome chirality

Wei Li, Shuo-Xing Dou, Peng-Ye Wang[*]

*Laboratory of Soft Matter Physics, Beijing National Laboratory for Condensed Matter Physics,*

*Institute of Physics, Chinese Academy of Sciences, Beijing 100080, China*

**Abstract**

In eukaryote nucleosome, DNA wraps around a histone octamer in a left-handed way. We study the process of chirality formation of nucleosome with Brownian dynamics simulation. We model the histone octamer with a quantitatively adjustable chirality: left-handed, right-handed or non-chiral, and simulate the dynamical wrapping process of a DNA molecule on it. We find that the chirality of a nucleosome formed is strongly dependent on that of the histone octamer, and different chiralities of the histone octamer induce its different rotation directions in the wrapping process of DNA. In addition, a very weak chirality of the histone octamer is quite enough for sustaining the correct chirality of the nucleosome formed. We also show that the chirality of a nucleosome may be broken at elevated temperature.

[*] Corresponding author. E-mail address: pywang@aphy.iphy.ac.cn

## 1. Introduction

In eukaryotes, a nucleosome is the basic structural element of chromatins. Each nucleosome contains a histone octamer and 146 bp of DNA. The histone octamer is composed of four pairs of histone H2A, H2B, H3 and H4 molecules. In natural nucleosomes DNA wraps around a histone octamer in about two turns with a left-handed chirality. The histone octamer's structure has been solved by x-ray crystallography (Arents et al., 1991; Luger et al., 1997). It reveals a tripartite assembly of the histone octamer in which a centrally located (H3-H4)$_2$ tetramer is flanked by two H2A-H2B dimmers, and the eight histone molecules form a left-handed protein superhelix (Widom, 1989; Hamiche et al., 1996; Ramakrishnan, 1997). The histones are put together in such a way that they define a left-handed wrapping path for the DNA via grooves, ridges and binding sites. With the development of single molecule manipulation methods, some works have been done on stretching DNA molecules and chromatin fibers (Katritch et al., 2000; Cui and Bustamante, 2000; Bennink et al., 2002; Brower-Toland et al., 2002). These works are helpful for further understanding the structure and function of chromatins. Meanwhile, some theoretical works have been done on the dynamical properties of DNA and chromatin (Noguchi et al., 2000; Sakaue et al., 2001; Kulić and Schiessel, 2004; Sakaue and Löwen, 2004). We have previously studied the process of nucleosome formation and that of disruption under stretching with Brownian dynamics simulation (Li et al., 2003, 2004). In these theoretical works, a simple spherical ball or a cylinder was used to model the histone octamer. Therefore, the chirality of the histone octamer was neglected. In the present paper, we present a new model for the histone octamer with a quantitatively adjustable chirality: left-handed, right-handed or non-chiral. Rather than using a single particle, we use a group of balls to model the histone octamer. The natural nucleosome feature that histones are put together to form a left-handed wrapping path for DNA can be included. The wrapping process of a DNA molecule, described as a wormlike chain, around the histone octamer is studied with Brownian dynamics simulation. With this new modeling, we can quantitatively analyze the chirality of the nucleosome formed and may obtain clues to the understanding of why DNA wraps around histone octamer only in the left-handed way in natural nucleosomes. In our model, the linker histone H1 (H5) is neglected for previous experiments have shown that it is not essential for nucleosomal array to assemble into higher-order structures (Shen et al., 1995; Ohsumi et al., 1993; Dasso et al., 1994).

## 2. Model

As in the previous works (Noguchi et al., 2000; Sakaue et al., 2001; Li et al., 2003, 2004; Sakaue and Löwen, 2004), the DNA molecule is modeled as a semi-flexible homopolymer chain (wormlike chain). Through out this work, we use 40 spheres connected by bonds for the homopolymer chain.

Rather than using a simple spherical ball or a cylinder as in the previous works (Noguchi et al., 2000; Sakaue et al., 2001; Li et al., 2003, 2004; Sakaue and Löwen, 2004), the histone octamer is presently modeled with a large spherical ball surrounded by five small spherical balls. All these balls are put together to form a histone octamer. The positions of the small balls determine the chirality of the histone octamer, as shown in Fig. 1.

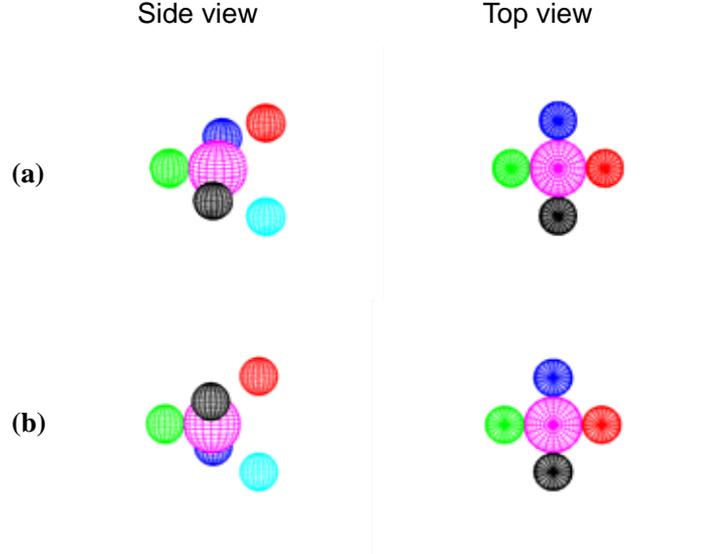

Fig. 1. Models for a histone octamer. (a) left-handed; (b)right-handed.

The interactions in this system are considered as follow.

The self-avoiding effect of DNA chain is considered by using the repulsive part of the Morse potential,

$$U_{m,rep} = \varepsilon_m k_B T \sum \exp\{-\alpha_m (r_{i,j} - \sigma_m)\}, \quad (1)$$

where $\varepsilon_m = 0.2$ and $\alpha_m = 2.4$. We set the Boltzmann constant $k_B$ to unity and choose 298 K for the temperature $T$. $r_{i,j}$ is the distance between the i*th* and j*th* spheres of the DNA chain. $\sigma_m$ is the minimum distance between to spheres of the DNA chain which gives the self-avoiding separation. Thus the diameter of one sphere of the DNA chain is $\sigma_m$. We use $\sigma_m$ as the length unit.

The bonds between neighboring spheres of the DNA chain are considered through harmonic bonding potential,

$$U_{bond} = \frac{k k_B T}{2\sigma_m^2} \sum \left(|\vec{r}_i - \vec{r}_{i+1}| - \sigma_m\right)^2, \quad (2)$$

where $k = 400$, $\vec{r}_i$ and $\vec{r}_{i+1}$ are the location vectors of the i*th* and (i+1)*th* spheres of the DNA chain.

The DNA chain stiffness is modeled by using the bending potential,

$$U_{bend} = \kappa k_B T \sum \left(1 - \frac{(\vec{r}_{i-1} - \vec{r}_i) \cdot (\vec{r}_i - \vec{r}_{i+1})}{\sigma_m^2}\right), \quad (3)$$

where we choose $\kappa = 5$ so that the DNA's persistence length in our model is consistent with the actual value, ~50 nm at physiological salt concentration of 0.1 M (Kunze and Netz, 2000; Vologodskaia and Vologodskii, 2002).

The interaction between the DNA chain and a histone octamer is simulated with the Morse potential,

$$U_M = \eta \varepsilon k_B T \sum [\exp\{-2\alpha(r_i - \sigma)\} - 2\exp\{-\alpha(r_i - \sigma)\}], \tag{4a}$$

$$U_M^{'} = \eta' \varepsilon k_B T \sum [\exp\{-2\alpha(r_i' - \sigma')\} - 2\exp\{-\alpha(r_i' - \sigma')\}], \tag{4b}$$

where $\eta = 0.9$, $\eta' = 0.02$, $\varepsilon = 6$, $\alpha = 6$, $\sigma = 2.0\sigma_m$ and $\sigma' = 1.0\sigma_m$. $U_M$ ($U_M^{'}$) is the potential of interaction between DNA and the big (small) histone ball. $r_i$ ($r_i'$) is the distance between the $i$th sphere of the DNA chain and the big (small) histone ball. We choose the parameter values for the balls so that the relative sizes of the modeled histone octamer and DNA are the same as those in a natural nucleosome.

The overdamped Langevin equations are used to describe the motion of each sphere of the DNA chain and the histone octamer,

$$-\gamma_m \frac{d\vec{r}_i}{dt} + \vec{R}_{m,t}(t) - \frac{\partial U}{\partial \vec{r}_i} = 0, \tag{5a}$$

$$-\gamma_M \frac{d\vec{R}_j}{dt} + \vec{R}_{M,t}(t) - \frac{\partial U}{\partial \vec{R}_j} = 0, \tag{5b}$$

where $\gamma_m$ and $\gamma_M$ are the friction constants of a sphere and a histone octamer, respectively. They are calculated according to Stokes law. $\vec{R}_{m,i}(t)$ and $\vec{R}_{M,j}(t)$ are the Gaussian white noises which obey the fluctuation-dissipation theorem,

$$<\vec{R}_{m,i}(t)> = 0, \quad <\vec{R}_{m,i}(t) \cdot \vec{R}_{m,j}(t')> = 6k_B T \gamma_m \delta_{i,j} \delta(t-t'), \tag{6a}$$

$$<\vec{R}_{M,j}(t)> = 0, \quad <\vec{R}_{M,i}(t) \cdot \vec{R}_{M,j}(t')> = 6k_B T \gamma_M \delta_{i,j} \delta(t-t'). \tag{6b}$$

The total internal energy $U$ consists of six terms:

$$U = U_{m,rep} + U_{bond} + U_{bend} + U_M + U_M^{'}. \tag{7}$$

We perform the dynamics of this system using a stochastic Runge-Kutta algorithm (white noise) (Honeycutt, 1992). We choose $k_B T$ as the unit energy, $\sigma_m$ as the unit length, and $\gamma_m \sigma_m / \sqrt{T}$ as the unit time step in our simulation.

We describe the rotation of histone octamer by the following equation,

$$\frac{d\vec{L}}{dt} = \vec{M}, \tag{8}$$

where $\vec{L}$ is the moment of momentum of the histone octamer and $\vec{M}$ is its moment of force. $\vec{L}$ is obtained by summing the moments of momentum of all the small histone balls. Here, we choose the center of the big ball as the instant center of the histone octamer's rotation.

## 3. Results

Firstly, we simulate the interaction between a DNA chain and a left-handed histone octamer. The dynamics of the interaction process is shown in detail in Fig. 2. It can be seen that DNA wraps around the histone octamer in about two turns in the left-handed way at last. And a left-handed nucleosome is formed. To show how DNA wraps around the histone octamer in detail, we zoom in the snapshot of the wrapped part of the histone octamer in Fig.2(g), as shown in Fig.3. Here, for clarity, the DNA chain is represented only by its backbone. From Fig.3 we can see that the DNA chain follows the chirality of the histone octamer in forming the left-handed nucleosome.

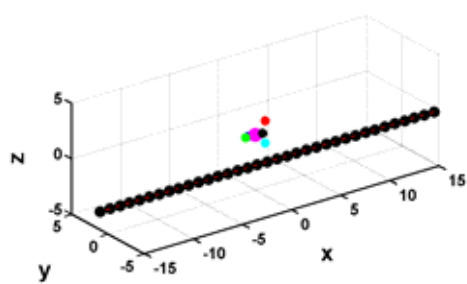

(a)

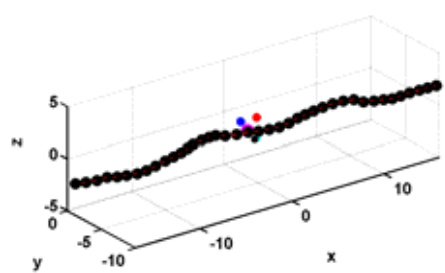

(b)

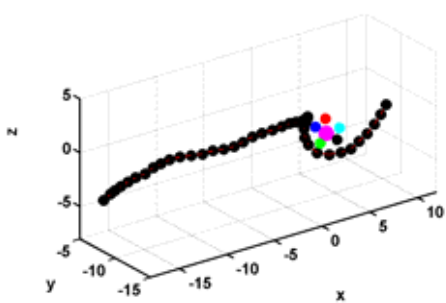

(c)

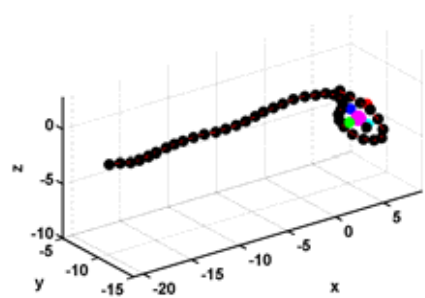

(d)

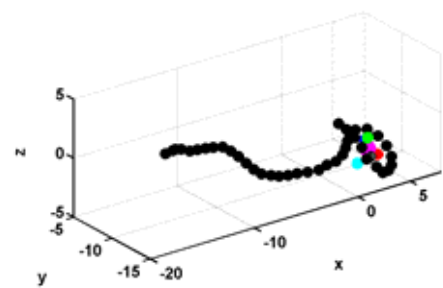

(e)

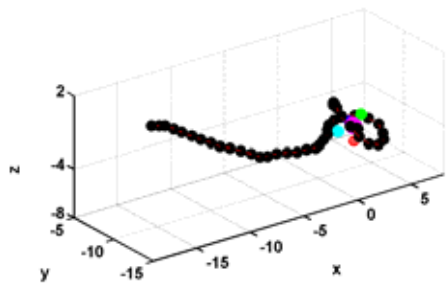

(f)

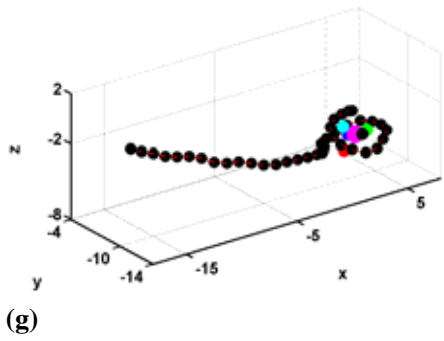 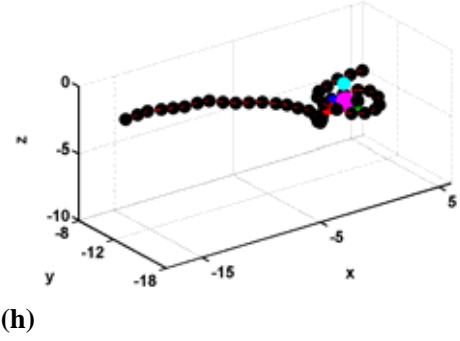

(g)          (h)

Fig. 2. Snapshots of the dynamics process of interaction between the DNA chain and a left-handed histone octamer. The simulation time $t$ is: (a) 0; (b) 0.88; (c) 61.6; (d) 70.4; (e) 88; (f) 96.8; (g) 105.6; (h) 114.4.

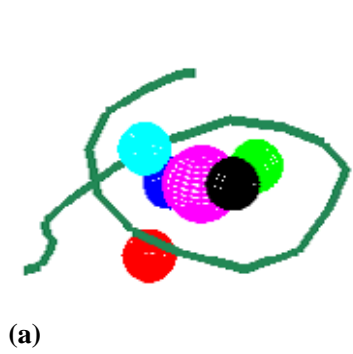 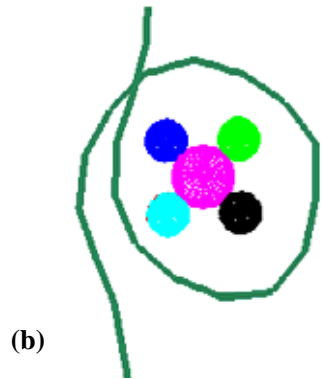

(a)          (b)

Fig. 3 Close-ups of the wrapped part of the histone octamer in Fig. 2(g). For clarity, only the backbone of the DNA chain is shown. (a) side view, (b) top view.

We show the snapshots of the histone octamer's movement in Fig. 4 corresponding to the pictures in Fig. 2. We can see that during this interaction process, the histone octamer also rotates, essentially in the clockwise direction.

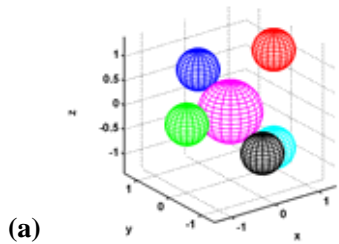 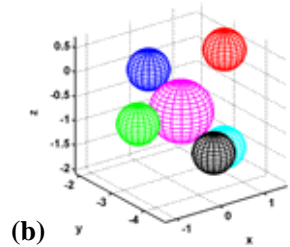

(a)          (b)

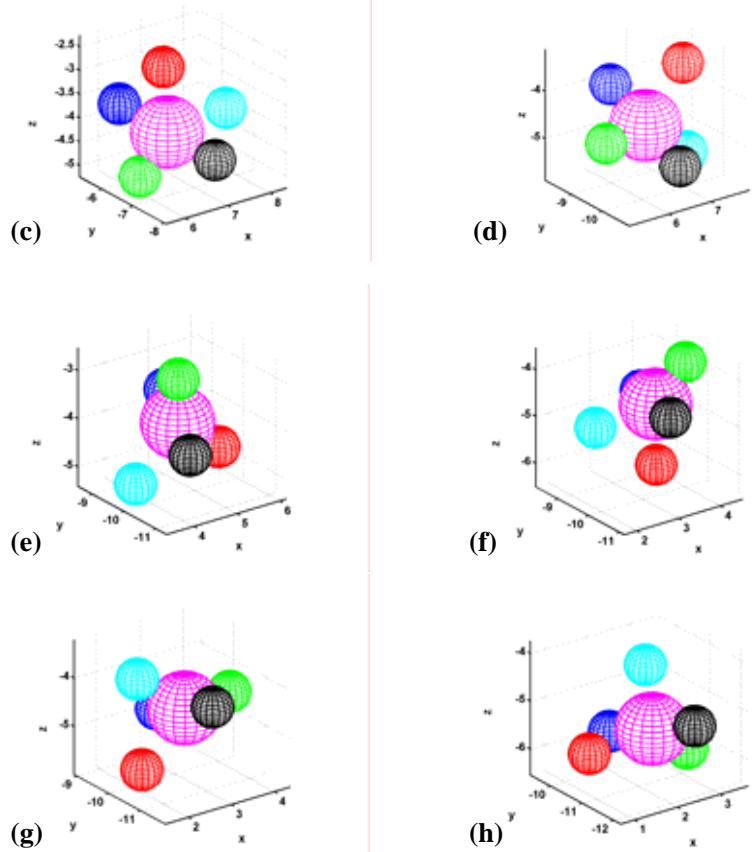

Fig. 4.　Snapshots of the movement of the left-handed histone octamer at times corresponding to that in Fig. 2.

Secondly, we simulate the interaction between the DNA chain and a right-handed histone octamer (for comparison). The results show that DNA chain wraps the histone octamer also in about two turns in the right-handed way at last. A right-handed nucleosome is formed. The histone octamer rotates in the anti-clockwise direction. (Snapshots are not shown here).

In our previous study (Li et al., 2004), we proposed a model to interpret the interaction between DNA and a histone octamer. We thought that the rotation of the histone octamer helped the DNA chain to wrap around it. In another recent work (Kulić and Schiessel, 2004), the rotation of the histone octamer was also considered. In the present paper, we make a modified model for the histone octamer so that we are able to see the formation of the nucleosome chirality. From the simulation results, we present a model for the interactions between DNA and a left-handed histone octamer to form a natural nucleosome (Fig. 5(a)). In Fig. 5(b) we plot right-handed histone octamer for comparison. That is, during the wrapping process, the histone octamer rotates, and the DNA chain follows the rotation while its two ends remain in their relative positions with respect to the histone octamer.

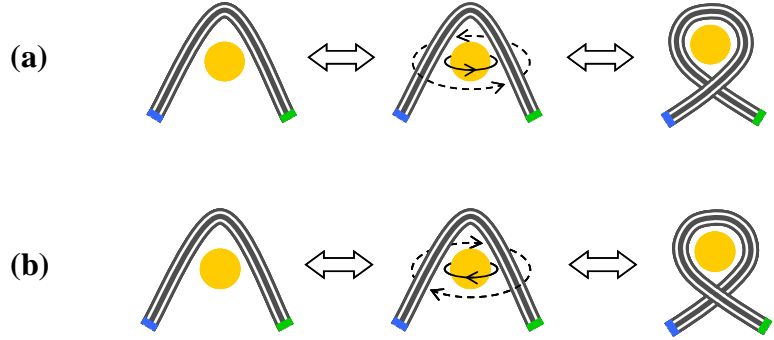

Fig.5. Model for interactions between DNA and a histone octamer. (a) left-handed histone octamer (natural case); (b) right-handed histone octamer (for comparison). The histone octamer is represented simply by a yellow ball.

From our simulation, we can see that in the process of interaction between DNA and a histone octamer, the histone octamer rotates with a relative rotation direction determined by its chirality. And the chirality of the nucleosome formed depends on the histone octamer's relative rotation direction. That is to say, the chirality of the nucleosomes formed is determined by the chirality of the histone octamer. In nature, only left-handed histone octamer exists, thus there is only left-handed nucleosome. For more than one histone octamer, it can be expected that in the process of interaction between a DNA and histone octamers, all histone octamers should rotate in the same direction relative to the DNA chain.

## 4. Discussion

We have studied the rotation of a histone octamer during its interaction with DNA. We find that the rotation of a histone octamer helps DNA wrap around it and the chirality of the nucleosome formed depends on the direction of the histone octamer's relative rotation which is, in turn, determined by the chirality of the histone octamer. Thus eventually it is the structure of the histone octamer that determines the chirality of a nucleosome. But how much does the chirality of nucleosome depend on the histone octamer's structure? In another word, how much is the histone octamer's chirality needed to sustain the correct chirality of the nucleosome? These are interesting questions that need to be answered.

We first make a comparison of our models for left-handed and right-handed histone octamers. As can be seen in Fig. 6, the only difference between the two histone octamers is in the different $z$ coordinates of the 2$^{nd}$ (blue) and 4$^{th}$ (black) small balls, $z_2$ and $z_4$. In the case of left-handed histone octamer, the $z$ coordinate of the black (blue) ball is lower (higher) than that of the central big ball. In the case of right-handed histone octamer, the reverse is true. Thus it's very easy for us to change the chirality of a histone octamer from one to another: we just need to vary $z_2$ and $z_4$. When the two balls have the same $z$ coordinate as that of the central big ball, i.e., $z_2 = z_4 = 0$, then the histone octamer has no chirality.

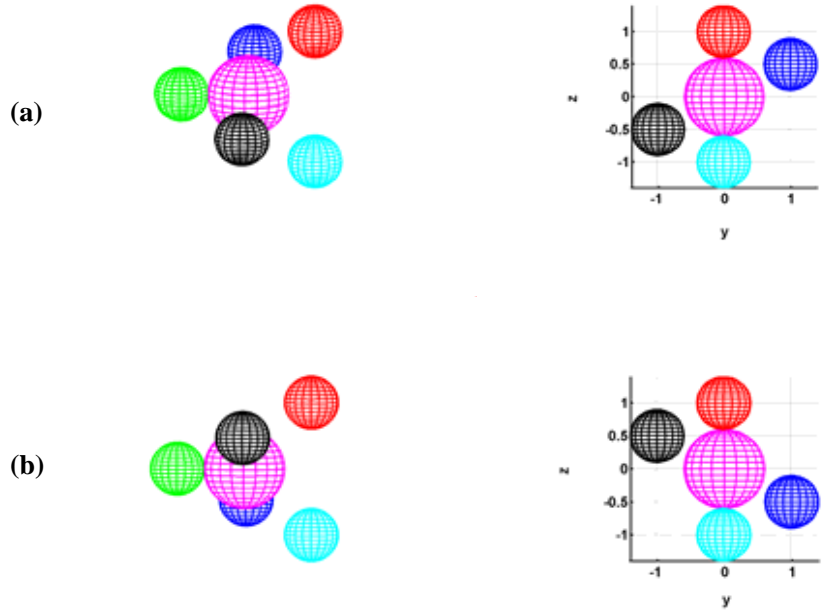

Fig. 6. The geometrical structures of our histone octamer models (left) and their side-views (right). (a) left-handed histone octamer; (b) right-handed histone octamer.

We change the $z$ coordinates of the 2$^{nd}$ and 4$^{th}$ small balls step by step and thus vary the chirality of the histone octamer accordingly. After each change, we repeat the simulation of the interaction between DNA and the histone octamer, and check the stability (or repeatability) of the chirality of the nucleosome formed. All the simulation results are shown Fig. 7. From Fig. 7, it can be seen clearly that as long as the distance between the 2$^{nd}$ and 4$^{th}$ small balls, $|z_2 - z_4|$, is larger than $\sim 5 \times 10^{-4}$, the chirality of the nucleosome formed is always the same as that of the histone octamer. And only when this distance is as small as $< 5 \times 10^{-4}$, nucleosomes of both chiralities start to appear. This means that a weak chirality of the histone octamer is quite enough for sustaining the correct (or stable) chirality of the nucleosome formed. When $z_2 = z_4 = 0$, the histone octamer has no chirality. Right-handed or left-handed nucleosomes occur with similar probabilities.

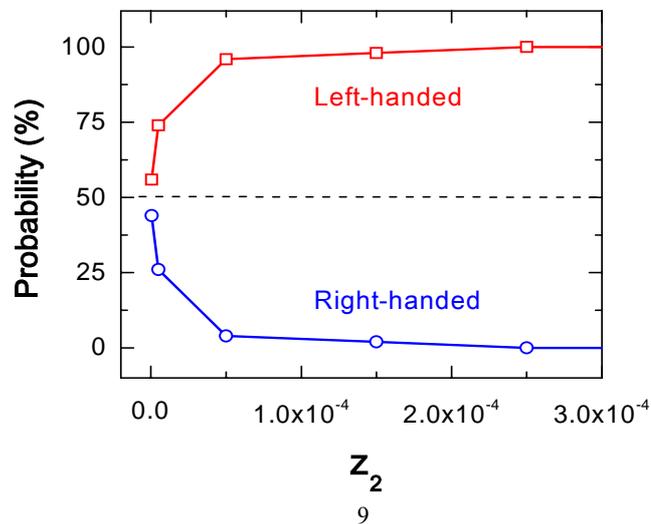

Fig. 7. Probabilities of occurrence of left-handed and right-handed nucleosomes versus the *z* coordinate of the 2nd small ball, $z_2$. To obtain each pair of data points at a given $z_2$, the simulation is repeated by at least 50 times. Note that $z_4$ is always taken as $z_4 = -z_2$.

Finally, we check the effect of temperature on the chirality of nucleosome. We choose a stable left-handed nucleosome structure and change the environment temperature. For each temperature change, we repeat the simulation to check the stability of the nucleosome structure. We find that with the increase of temperature, the right-handed nucleosome appears with increasing probability. The result is shown in Fig. 8. Note that, in Fig. 8, the high temperature limit is 400 K which is already beyond the reasonable value for nucleosomes, since even DNA molecules will be denatured at about 368 K. But these results demonstrate that the chirality of a nucleosome is very stable against temperature variation of the environment. We find that if the temperature is further increased the nucleosome structure will be completely destroyed (data not shown).

In addition, we know that the chirality of DNA is mainly right-handed. But whether the chirality of DNA has an effect on the chirality of nucleosome is still unknown. We think that the chirality of DNA may have little effect on the chirality of nucleosome. In our future work, we will consider the chirality of DNA and check its effect on the chirality of nucleosome.

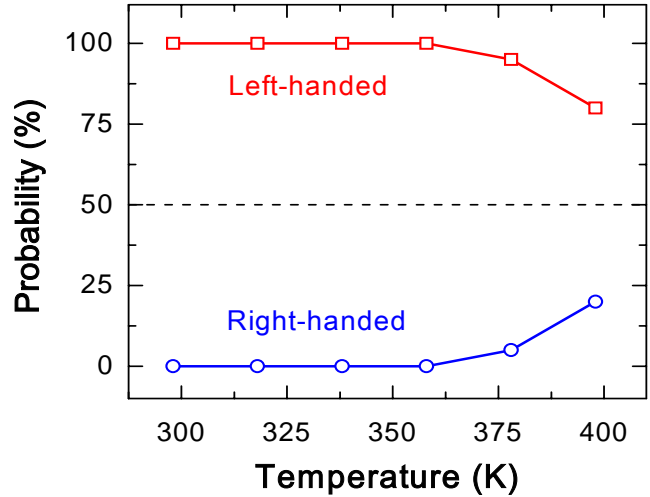

Fig. 8. Probabilities of occurrence of left-handed and right-handed nucleosomes versus temperature. The chirality of the histone octamer is left-handed with $z_2 = -z_4 = 2.5 \times 10^{-4}$. To obtain each pair of data points at a given temperature, the simulation is repeated by at least 20 times.

## 5. Summary

We build up a new model for histone octamers to investigate the effect of histone octamer's structure on the chirality of nucleosome formed. From our simulation results, we find that the chirality of the nucleosome is determined by that of the histone octamer. In nature, only left-handed histone octamers exist, so there are only left-handed nucleosomes. We also find that a weak chirality of the histone octamer is enough for the nucleosome to get the right structure (chirality). In addition, the temperature plays important role in the formation of nucleosome structure. In a normal range of temperature, the nucleosome can get the right structure, but if the temperature is increased, the chirality of the nucleosome formed may be broken.

From the detailed dynamic interaction processes of a histone octamer with DNA, we find that the histone octamer is rotating when interacting with DNA. We thus proposed a model to explain the interaction between histone and DNA. In this model, the histone octamer will rotate in a relative direction (with respect to DNA) that is determined by its chirality, and DNA will follow the rotation in wrapping around the histone octamer. Thus it is expected that when a long DNA chain interact with multiple histone octamers in assembling into a chromatin, the DNA chain will locally follow the rotations of all the histone octamers, and then multiple nucleosomes of the same chirality will be formed.

**Acknowledgements**

This research was supported by National Natural Science Foundation of China, and the Innovation Project of the Chinese Academy of Sciences.